\documentclass[12pt]{iopart}

  \expandafter\let\csname equation*\endcsname\relax
  \expandafter\let\csname endequation*\endcsname\relax
\usepackage{amsmath}
\usepackage{amssymb}
\usepackage{upgreek}
\usepackage{float}
\usepackage{graphicx}
\usepackage{wrapfig}
\usepackage{color}
\usepackage{epstopdf}

\usepackage{iopams}

\newcommand{\vect}[1]{{\boldsymbol{#1}}}

\makeatletter

\begin{document}

\title[Momentum independent pseudogap in DOS and Raman spectroscopy.]{Signatures of a momentum independent pseudogap in the electronic density of states and Raman spectroscopy of the underdoped cuprates.}

\author{J. P. F. LeBlanc$^{1}$}

\address{$^1$Max-Planck-Institute for the Physics of Complex Systems, 01187 Dresden, Germany} 
\ead{jpfleblanc@gmail.com}


\date{\today}
\begin{abstract}
We propose a hybridization phenomenology to describe the pseudogap state of the underdoped cuprates. We show how a momentum independent pseudogap opens asymmetrically from the Fermi-surface but symmetric to the zeroes of the hybridized bonding dispersion, which results in false d-wave characteristics of the pseudogap at the Fermi level.    By comparing against a d-wave form factor we illustrate the difficulty in identifying a momentum independent order in momentum averaged quantities such as the electronic Raman response.  We identify  a suppression in the single-particle density of states which produces a hump feature which should be observable experimentally in tunnelling $dI/dV$ spectra and distinguishes the s-wave and d-wave ordering scenarios.
\end{abstract}

\maketitle

\section{Introduction}

There remains no consensus as to the mechanism behind the anomalous normal state pseudogap, or its connection, if any, to superconductivity in the underdoped cuprates.\cite{timusk:1999,kohsaka:2012,  valla:2006}
However, one  common viewpoint is that the pseudogap stems from strong correlation physics, for which the simplest model is the 2D Hubbard model.
While it may be that the answer to the pseudogap problem is to compute the exact electronic correlations throughout the phase diagram,\cite{gull:2012, leblanc:2013} what the community at large strives for is instead the identification of a simple mechanism, or solvable effective Hamiltonian which contains the essence of the correlations responsible for the pseudogap.  There have been many such proposals, such as, to name a few: loop currents,\cite{varma:2010} formation of spin singlets,\cite{lee:2006,  ogata:2008, yrz:2006} charge or spin density waves (d-density wave),\cite{chakravarty:2001} the spin-fermion model,\cite{abanov:2000} quasi 1D spin liquids\cite{tsvelik:2007} and more recent charge density fields emerging from fluctuating antiferromagnetic orders.\cite{sachdev:2013, hayward:2013}
 While each model is interesting in its own regard, it is not clear which, if any, is describing the physics which gives rise to the pseudogap.

Distinguishing between various phenomenologies requires direct experimental evidence.  A monumental experimental observation was the work of Hashimoto et al.\cite{hashimoto:2010} who identified that the pseudogap is particle-hole asymmetric.  
  This has been verified through the subsequent observation   of new Bogoliubov quasiparticle bands which appear below the superconducting transition temperature $T_c$ only in the presence of a particle-hole asymmetric pseudogap.\cite{he:2011, leblanc:2011:bqp} 
If the pseudogap is not particle-hole symmetric, then any representation of the pseudogap along the Fermi momenta will not accurately depict the momentum dependence of the pseudogap order.  Such a conclusion has been hinted at by recent experimental and computational work.\cite{sakai:2010,sakai:2013,sakai:2013:2,sacuto:2012,sacuto:2013}

The goals of this paper are two-fold: 1) show that there exists at least one model where the inclusion of a momentum independent (s-wave) order-parameter results in an apparent d-wave pseudogap at the Fermi level. 2) Identify signatures of a momentum independent gap in momentum averaged observable quantities. 
To accomplish these goals, we employ a hybridization phenomenology consistent with one of the most successful models of the pseudogap phase.\cite{yrz:2006, rice:2012, comin:2014}
We examine momentum resolved quantities, such as spectral density and energy dispersion, as well as momentum averaged quantities.  We present results for the Raman B$_{1g}$ and B$_{2g}$ spectra in order to identify the features of the onset of a pseudogap order with either $s-$ or $d-$wave symmetry.  We also compute the single particle density of states in which we identify the most clear signatures of a momentum independent pseudogap from a momentum averaged quantity.

In Section \ref{sec:theory} we will build the relevant theoretical framework.  Section \ref{sec:resolved} will contain results of the phenomenological band structure, Section \ref{sec:averaged} will focus on momentum averaged quantities, and Section \ref{sec:conc} will contain a brief summary.


\section{Theory}\label{sec:theory}
\subsection{Hybridization Phenomenology}\label{sec:phenom}

Within a BCS (Bardeen-Cooper-Schrieffer) theory the superconducting gap  originates at the intersection of two  field operators, namely the fermionic electron (hole) creation (annihilation) operators.  In the standard problem the dispersion takes the simplified form $E_{\rm sc}(\vect{k})=\pm\sqrt{\epsilon_{\vect{k}}^2+\Delta_{\rm sc}^2(\vect{k})}$,  which represents a superconducting gap, $\Delta_{\rm sc}(\vect{k})$, opening at the Fermi level of the electronic dispersion, $\epsilon_{\vect{k}}$. 
Therefore, the energy dispersion at a point along the Fermi momentum, $\vect{k}_F$, reduces to $E_{\rm sc}(\vect{k}_F)=\pm \Delta_{\rm sc}(\vect{k}_F)$.  We will see that this collapse at $\vect{k}=\vect{k}_F$ is not relevant for a particle-hole asymmetric pseudogap, and identify instead the momenta about which the dispersion of the pseudogap state collapses.

To accomplish this, we consider a general hybridization phenomenology of electron operators, $c_{\vect{k}\sigma}$, coupling with an excitation defined by the operator, $D_{\vect{k}\sigma}$, and we specify the dispersions of these particles throughout the Brillouin zone as $\epsilon_{\vect{k}}^{c(D)}$ respectively.  These two operators will share a constant energy contour in momentum space where $\epsilon_{\vect{k}}^c=\epsilon_{\vect{k}}^D$ along which hybridization can occur, reducing the overall energy of the system, forming a gap.  
The hybridization scenario follows as:\cite{imada:2013, yamaji:2011}

\begin{align}
\mathcal{H} = \sum\limits_{\vect{k}\sigma}\epsilon_{\vect{k}}^c c_{\vect{k}\sigma}^\dagger c_{\vect{k}\sigma} + \epsilon_{\vect{k}}^D D_{\vect{k}\sigma}^\dagger D_{\vect{k}\sigma} - \lambda \left(c_{\vect{k}\sigma}^\dagger D_{\vect{k}\sigma}+ D_{\vect{k}\sigma}^\dagger c_{\vect{k}\sigma}\right).
\end{align}

The choice of $D_{\vect{k}\sigma}$ operators may differ depending on the model.\cite{smith:2010} For example in most charge, spin or d-density wave models the operator  $D_{\vect{k}\sigma}=c^\dagger_{\vect{k}+\vect{Q} \sigma}$, an electron creation operator translated through a $\vect{Q}$ scattering momentum.
To continue, the Hamiltonian can be rewritten as
\begin{align}\label{eqn:hamiltonian}
\mathcal{H}=\sum\limits_{\vect{k}\sigma} \Psi^\dagger(\vect{k}) \begin{pmatrix} \epsilon_{\vect{k}}^c & -\lambda \\ -\lambda^* & \epsilon_{\vect{k}}^D   \end{pmatrix}    \Psi(\vect{k}) 
\end{align}
where $\Psi^\dagger(\vect{k}) = \begin{pmatrix}
c_{\vect{k}\sigma}^\dagger & D_{\vect{k}\sigma}^\dagger
\end{pmatrix}  $.  The problem is then diagonalized by defining $\epsilon_{\vect{k}}^-=\frac{\epsilon_{\vect{k}}^c-\epsilon_{\vect{k}}^D}{2}$  and $\epsilon_{\vect{k}}^+=\frac{\epsilon_{\vect{k}}^c+\epsilon_{\vect{k}}^D}{2}$, bonding and anti-bonding bands respectively. 
The result is a two-band system with energies given by $E_{\vect{k}}^\pm = \epsilon^+_{\vect{k}} \pm E_{\vect{k}}$, where $E_{\vect{k}}= \sqrt{{\epsilon_{\vect{k}}^-}^2 +\lambda^2  }$. 
From the Hamiltonian of Eq.~(\ref{eqn:hamiltonian}) we can write the Green's function using $G(\vect{k},i\omega_n) ^{-1} = i\omega_n - \mathcal{H}$.  While the Hamiltonian and Green's function are matrices, we consider only the coherent electronic element,  $\left\langle  c_{\vect{k}\sigma}^\dagger c_{\vect{k}\sigma} \right\rangle$, of the Green's function which can be written as 
\begin{eqnarray}
G(\boldsymbol{k},\omega)=
\sum_{\alpha=\pm}  \frac{{|W^{\alpha}_{\boldsymbol{k}}|^2	}}{\omega-E^{\alpha}_{\boldsymbol{k}}+ i\Gamma},
\end{eqnarray}
 where $|W_{\vect{k}}^\pm|^2= \frac{1}{2}(1 \pm \frac{\epsilon^-_{\vect{k}}}{E_{\vect{k}}})$ and are assumed to be real.
 Off diagonal elements can be treated as well\cite{valenzuela:2005}, however we will see in the following section that the model presented there neglects off diagonal contributions.

The resulting spectral density, $A(\vect{k},\omega)= -2 {\rm Im} G(\vect{k}, \omega )$, will show a hybridization gap, $\lambda$, which opens along the momenta corresponding to zeros of the bonding dispersion.  If we define the momenta where this occurs to be $\vect{k}_B$, such that $\epsilon_{\vect{k}_B}^-=0$, then the energy dispersion reduces to $E_{\vect{k}_B}^\pm=\epsilon_{\vect{k}_B}^+ \pm \lambda$.
The distinction between a gap opening at the Fermi momentum, $\vect{k}_F$, (as was the case for a superconducting gap) from a gap opening away from the Fermi momentum, in this case at $\vect{k}_B$, is the energy of the antibonding dispersion.
 We can therefore obtain a pseudogap in the hybridization scenario without resorting to a complicated momentum dependence of the order parameter, $\lambda$. Instead, the momentum dependence of the pseudogap has been moved off of $\lambda$, and is an element of the bonding and antibonding dispersions, $\epsilon_{\vect{k}}^-$ and $\epsilon_{\vect{k}}^+$ respectively, which contain elements of both the $c_{\vect{k}\sigma}$ and $D_{\vect{k}\sigma}$ fermion dispersions.  This will be central to understanding the s-wave picture of the pseudogap.
  

\subsection{Doping dependence through a Yang, Rice, Zhang Ansatz}
An optimal starting point for including a doping dependent phase diagram is the rather successful ansatz for the pseudogap self energy due to Yang, Rice and Zhang (YRZ)\cite{yrz:2006, yrz:2009, leblanc:2009, illes:2009, carbotte:2010:kag, leblanc:2010, ashby:2013} which was recently reviewed.\cite{rice:2012}  Recent work has reproduced this ansatz from a strong theoretical footing by mapping to a slave boson mean field theory  in which fermions are factored into spinon and holon constituent particles which can then bind together forming a gap at the particle boundary.\cite{james:2012}  This is therefore one of several phenomenologies which maps to the hybridization formulation of Sec.~\ref{sec:phenom}.

The spirit of the YRZ ansatz is to start from an overdoped Fermi liquid system and build in a mechanism for the suppression of electronic states at the Fermi level originating from Mott­ insulating physics as the system is underdoped.  What results is a simple non­interacting set of bands which contain the key elements needed to describe the basic qualitative features of experiments in the pseudogap state.\cite{rice:2012}  Naturally this approach has an intrinsic success related to low-­energy ‘hot­spot’ physics, which will break down under extreme underdoping and will never correctly capture effects due to scattering.
 The original YRZ ansatz, produced through comparison to calculations of 2 and 4 leg Hubbard ladders,\cite{konik:2006} enforced a d-wave symmetry to the pseudogap order parameter in the standard form $\Delta_{pg}^d(\vect{k})=0.5 \Delta_{pg}(\cos k_x - \cos k_y)$.  
Here we will compare this standard assumption of the d-wave pseudogap form factor to a momentum independent order parameter  which we will abbreviate throughout as $\Delta_{\rm pg}^s=\Delta_{\rm pg}$.  In both cases we define the pseudogap scale as $\Delta_{\rm pg}=A_0 \Delta_{\rm pg}^0(x)$  where $x$ is doping, $\Delta_{\rm pg}^0(x)$ is a doping dependent energy scale, and $A_0$ is a prefactor representing temperature dependence such that $A_0=1$ at $T=0$ and $A_0=0$ at the pseudogap critical temperature, $T^*$. \cite{sakai:2010,sakai:2013} 

The YRZ model can be phrased in terms of a hybridization process by taking the  electronic dispersion, $\epsilon_{\vect{k}}^c= - 2t(x)(\cos
k_xa  + \cos k_ya ) - 4t^{\prime}(x) \cos k_xa \cos k_ya - 2t''(x)(\cos
2k_xa  + \cos 2k_ya )-\mu_p$ and $\epsilon_{\vect{k}}^D =   2t(x)(\cos k_xa  + \cos k_ya)$.\cite{yrz:2009}  The electronic dispersion is the third
nearest-neighbor tight-binding energy from a Gutzwiller projected t-J Hamiltonian which acts on an RVB wavefunction.  This produces an effective tight binding dispersion which contains Gutzwiller renormalization factors, hopping amplitudes $t$-$t^\prime$-$t^{\prime\prime}$ and an electronic chemical potential, $\mu_p$, which sets the filling of the electronic dispersion.  
The $\epsilon_{\vect{k}}^D$ dispersion is fundamental to the YRZ ansatz and represents a particle which has a Fermi surface that is independent of doping, related to the half-filled Mott insulator, and which is perfectly susceptible to a vector nesting of $Q=(\pi,\pi)$ at the Fermi level.
To continue, we construct the coherent piece of the single particle Green's function which results in
\begin{eqnarray}
G(\boldsymbol{k},\omega,x)=
\sum_{\alpha=\pm}  \frac{{g_t(x) |W^{\alpha}_{\boldsymbol{k}}|^2	}}{\omega-E^{\alpha}_{\boldsymbol{k}}+ i\Gamma},
\label{eqn:sc}
\end{eqnarray}
where $g_t(x)= \frac{2x}{1+x}$.
 The energy dispersions and weighting factors now contain doping dependent hopping parameters, the details of which are located in \ref{sec:app:yrz}.  It is this doping dependent YRZ Green's function, Eq.~(\ref{eqn:sc}), which we employ in the following sections.
 

\section{Momentum Resolved Quantities}\label{sec:resolved}

\begin{figure*}
\centering
\includegraphics[width=.8\linewidth]{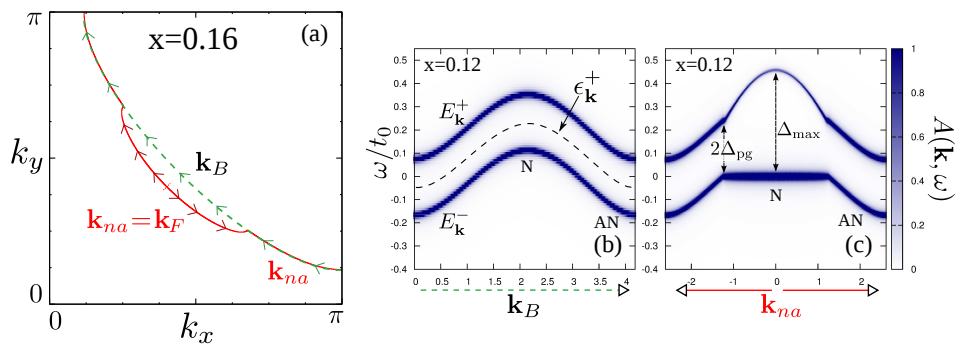}
 \caption{(Color Online) (a) An example at $x=0.16$ of the relevant $k$-space contours, $\vect{k}_B$ and $\vect{k}_{na}$, at the Fermi level. (b) Spectral density $A(\vect{k}_B, \omega, x=0.12)$ where $\vect{k}_B$ is the length along the green dashed curve of frame (a).  (c)  Spectral density $A(\vect{k}_{na}, \omega, x=0.12)$ for $\vect{k}_{na}$ along the nearest approach contour, the red curve of frame (a). All quantities were evaluated with $A_0=0.5$ using a momentum independent order, $\Delta_{\rm pg}^s$.}
\label{fig:nacont}
\end{figure*}

\subsection{Hybridized Spectral Density}

We explore the impact of a momentum independent pseudogap on perceived ARPES spectra.  In Fig.~\ref{fig:nacont}(a) we trace the momentum and energy dependence of the spectral density $A(\vect{k},\omega,x=0.16)$ along two contours in momentum space.  These contours are labelled $\vect{k}_{B}$ and $\vect{k}_{na}$.  The first contour, $\vect{k}_B$, shown as a green-dashed curve  represents the contour in momentum space along which the bonding dispersion, $\epsilon_{\vect{k}}^-$, is equal to zero.  The second contour, shown in red, is the contour of nearest-approach (or minimum gap locus), which in this phenomenology is equivalent to the zeros of the bonding dispersion, except where there is a Fermi surface near the nodal region. 
In Fig.~\ref{fig:nacont}(b) and (c) we show the spectral density along $\vect{k}_B$ and $\vect{k}_{na}$ respectively for an example case at $x=0.12$.  In Fig.~\ref{fig:nacont}(b) one can see the momentum independent pseudogap and the collapse of the hybridized dispersion, \emph{i.e.} $E_{\vect{k}_B}^\pm \to \epsilon_{\vect{k}_B}^+ \pm \Delta_{\rm pg}^0$.  The splitting of the band is not symmetric about the Fermi level, but is instead symmetric about the antibonding dispersion, $\epsilon_{\vect{k}}^+$, traced as the dashed line in Fig.~\ref{fig:nacont}(b).
If one was instead to follow the set of $\vect{k}_{na}$  the result obtained is Fig.~\ref{fig:nacont}(c).  Indeed, at negative $\omega$ these two cases are essentially identical save for the existence of a Fermi surface in the nodal region in Fig.~\ref{fig:nacont}(c).  What is apparent is that the gap at the antinodes in Fig.~\ref{fig:nacont}(c) is not caused by a momentum dependent pseudogap, but instead has a simpler explanation in the dispersive nature of the antibonding band, $\epsilon_{\vect{k}}^+$.

\begin{figure}
\centering
  \includegraphics[width=.6\linewidth]{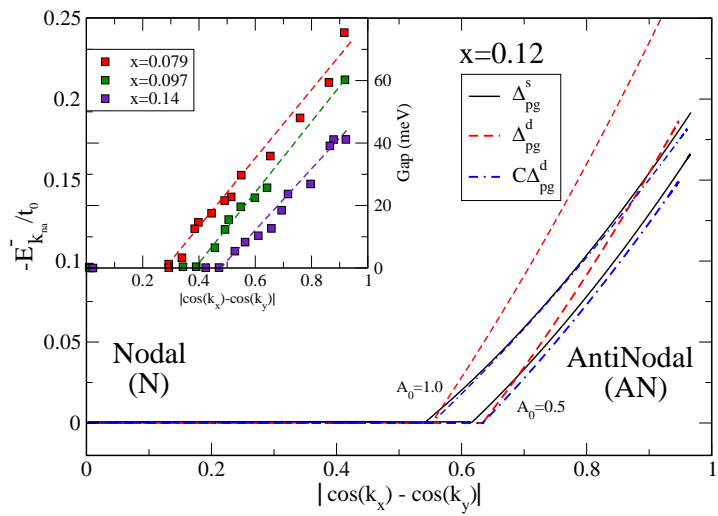}
    \caption{(Color Online) The dominant negative energy spectral peak with of a momentum independent order, $\Delta_{\rm pg}^s$ (solid lines), plotted as a function of the amplitude of the standard d-wave gap $\vert \cos(k_x) -\cos(k_y)  \vert$ compared with the standard $\Delta_{\rm pg}^d$ (dashed lines).  For illustrative purposes, an arbitrary scaling factor $\rm C$ can be applied to each $\Delta_{\rm pg}^d$ case to fit the $s-$wave order (blue double-dashed-dotted lines).  $E_{\vect{k}}^-$ is evaluated along $\vect{k}_{na}$ as the nearest approach contour is traversed from the nodal point (left) to the antinodal point (right).  Inset: Data on Bi-2212 from Vishik et al. Ref.~\cite{vishik:2012} for a range of doping, $x$, above the superconducting transition temperature.}
\label{fig:coskxy}
\end{figure}

To further emphasize this point, we show a common
 treatment of ARPES experimental spectra which is to plot the negative energy spectral peak, here $-E^-_{\vect{k}_{na}}$, as a function of the amplitude of the d-wave form as  done in Fig.~\ref{fig:coskxy}.  The momentum independent gaps, $\Delta_{\rm pg}^s$ (solid lines), produce what is an apparent d-wave pseudogap at the Fermi level in the antinodal region.  This is purely caused by the antibonding dispersion, $\epsilon_{\vect{k}}^+$, which we can see from Fig.~\ref{fig:nacont}(b) has a cosine-like behavior.  
The inclusion of a d-wave form factor, $\Delta_{\rm pg}^d$ (red dashed), only acts to increase the apparent slope of the curve for the same parameter values and can be absorbed into the scaling factor of $t_0$.  The inclusion of such a scaling factor $t_0 \to t_0/C$ is also shown in Fig.~\ref{fig:coskxy} (blue double-dashed-dotted lines), illustrating that the $s-$ and $d-$wave cases are qualitatively identical.  
Increasing $\Delta_{\rm pg}$ or decreasing the doping, $x$, results in a shift of the apparent gap onset to a lower value, which is a generic feature of models with Fermi arcs or pockets.
 Comparison to experimental data in the absence of superconductivity,  shown in the inset of Fig.~\ref{fig:nacont}, reveals that both $\Delta_{\rm pg}^s$ and $\Delta_{\rm pg}^d$ are possible candidates to explain the momentum dependence of ARPES spectra along the minimum gap locus since neither deviate strongly from the experimentally suggested linear trend. 
   The details of such a calculation in the superconducting state have been presented previously which adds only a nodal $d-$wave component.\cite{leblanc:2010, valenzuela:2007}

To summarize, within this phenomenology  a momentum independent gap opens along the bonding dispersion which results in symmetric bands relative to the energy of the antibonding dispersion and this will appear as a false d-wave pseudogap along the Fermi surface and at negative energies where ARPES can probe.  As a result, both the $\Delta_{\rm pg}^s$ and $\Delta_{\rm pg}^d$ scenarios are consistent with ARPES observations.

\section{Impact on Momentum Averaged Quantities}\label{sec:averaged}
\subsection{$B_{1g}$ and $B_{2g}$ Raman Spectra}
Raman spectroscopy has been an influential technique for examining the underdoped cuprates due to an ability to control which region of momentum space is sampled.  For example, the B1g (antinodal) and B2g (nodal) Raman shifts have been used in the past to examine the normal state\cite{venturini:2002, chen:1997, nemetschek:1997} but also to separate the pseudogap and superconducting energy scales.\cite{tacon:2006}
In the normal state, experiments could be understood through the inclusion of an anisotropic insulating gap originating from a ‘hot­spot’ near the antinode with an angular $d-$­wave order parameter with a $\cos(2\phi)$ dependence in the Brillouin zone angle, $\phi$.  In the present formulation of a hybridization phenomenology, and given 
Given the difficulty distinguishing $\Delta_{\rm pg}^s$ and $\Delta_{\rm pg}^d$ in momentum resolved ARPES, it is not at all understood how the Raman spectra with $s-$ or $d-$wave pseudogap are distinct.  We are therefore looking to start with the simplest model system where a distinction between an $s­ $ or $d­-$wave structure of the pseudogap might be seen.  For this reason, we compute the Raman shift at the one-loop level and employ the effective mass approximation for the B1g and B2g polarization vertices, $\gamma_{\vect{k}}^{B_{1g}}$ and $\gamma_{\vect{k}}^{B_{2g}}$ respectively. 
We also fix the scattering rate to be constant and all other parameters to be the same between the $s-$­wave and $d-$­wave cases.  This should result in the clearest signatures of $s-$­ or $d-$­wave order that are possible from a momentum averaged Raman spectra.
 The calculation of Raman spectra follows from what are now standard treatments from Devereaux et al.\cite{devereaux:1996}.  The extension of such a treatment to include multiband systems \cite{boyd:2009} results primarily in extra transitions between bands (interband) in addition to the low energy Drude­like (intraband) transitions which are sensitive to the band structure at the Fermi level.  
 For what follows, a minimal set of relevant details can be found in \ref{sec:app:raman}.

\begin{figure}
\centering
  \includegraphics[width=.55\linewidth]{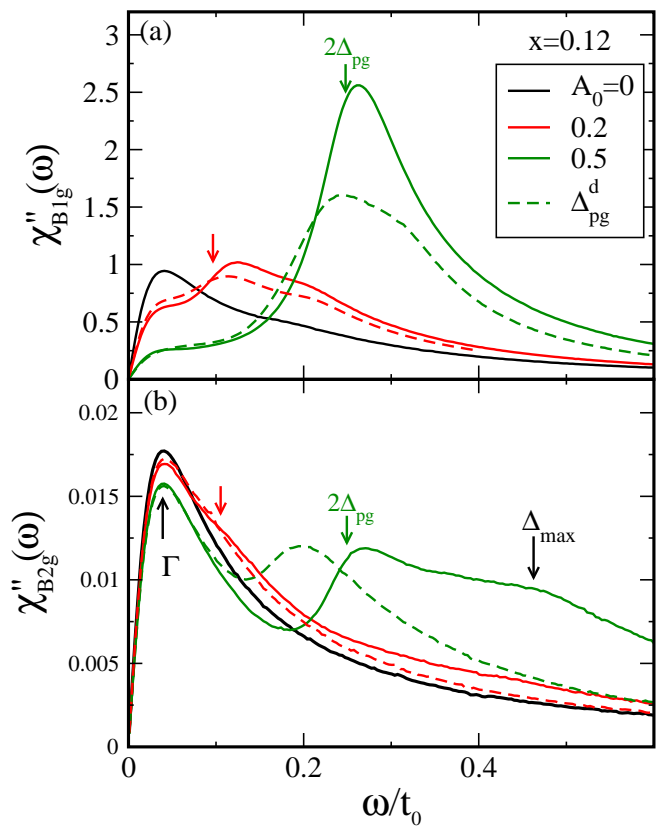}
    \caption{(Color Online) (a)The B1g Raman response, $\chi_{B1g}^{\prime \prime}(\omega)$, for a representative case of $x=0.12$.  (b)The corresponding B2g Raman response, $\chi_{B2g}^{\prime \prime}(\omega)$.  Calculations for both $\Delta_{\rm pg}^s$ (solid) and $\Delta_{\rm pg}^d$ (dashed) are shown.}
\label{fig:raman}
\end{figure}

A representative case at $x=0.12$ is shown in Fig.~\ref{fig:raman} employing a momentum independent order, $\Delta_{\rm pg}^s$ (solid lines).  Shown in black is the unhybridized case, where $\Delta_{\rm pg}^s=0$.  In that case, the band structure is a single band, and the Raman spectra represent only an intraband Drude-like response.  At the onset of $\Delta_{\rm pg}$, the single band breaks into two, $E_{\vect{k}}^\pm$.  The Raman spectra will then be comprised of a low energy intraband response and a higher energy interband transition, whose onset at $2\Delta_{\rm pg}$ is marked by coloured arrows above each curve.  The connection to a Drude response is important
because the low energy spectra is dominated by this response, which gives a peak in both $\chi^{\prime \prime}_{B1g}$ and $\chi^{\prime \prime}_{B2g}$ near $\omega=\Gamma$, where $\Gamma$ is the elastic scattering rate.  

As the pseudogap increases, the B$_{1g}$ spectra is extremely suppressed at low energy due to the erosion of Fermi surface which occurs in the antinodal region and is strongly sampled by the $\gamma_{\vect{k}}^{B1g}$ vertex.
To contrast, in the B2g spectra the slope in the limit of $\omega \to 0$ is determined by the scattering rate and is essentially unmodified for increasing pseudogap.  As a result there is only very weak suppression of the Drude response whose peak location does not shift with the value of $\Delta_{\rm pg}^s$.  It is important to point out that this suppression of the intraband response in either polarization contains no information regarding the scales of the pseudogap, since it contains only intraband ($E_{\vect{k}}^- \to E_{\vect{k}}^-$) transitions.  In the case of $\chi_{B2g}^{\prime \prime}$ this intraband suppression is weak due to the loss of states at the Fermi level, which occurs primarily in the antinodal region, and which is sampled only very weakly by the $\gamma_{\vect{k}}^{B_{2g}}$ polarization vertex.

At larger values of $\Delta_{\rm pg}$, both the $\chi^{\prime \prime}_{B1g}$ and $\chi^{\prime \prime}_{B2g}$ spectra exhibit clear interband transitions at the same frequency of $\omega=2\Delta_{\rm pg}^s$.  These corresponding peaks are the lowest energy feature which is evidence for an s-wave pseudogap.  
A final point of interest is that the highest energy scale would occur in the nodal region for the transition $\omega= \Delta_{\rm max}$ shown in Fig.~\ref{fig:nacont}(c).  This is the end of the interband transitions and is the least distinct indication of an ordering in the nodal region.
We show also for $A_0=0.2$ and $0.5$ the case with the $\Delta_{\rm pg}^d$ form factor (dashed lines).  What is evident is that the low energy Raman response with $\Delta_{\rm pg}^d$ is not qualitatively distinct from the momentum independent case in either the $\chi^{\prime \prime}_{B1g}$ or $\chi^{\prime \prime}_{B2g}$ spectra.  Both the $s-$ and $d-$wave cases show a similar low energy suppression. With $\Delta_{\rm pg}^d$ the interband transitions occur at lower energy in the nodal region due to having a smaller value of the gap at the edge of the Fermi pocket for the same input parameters, a previously understood effect.\cite{leblanc:2010}  Finally, the high energy feature at $\Delta_{max}$ does not occur in the d-wave case, since the gap would vanish in the nodal region, and therefore cannot result in the transition shown in Fig.~\ref{fig:nacont}(c).

To summarize, examining the loss of low energy Raman signal due to the onset of a pseudogap does not allow for a clear distinction between an $s-$ or $d-$wave pseudogap structure factor.  Instead, both cases present a similar loss of Fermi surface in the antinodal region which reduces the Drude response, and this does not appear to be an indication of a nodal pseudogap ordering.  A momentum independent pseudogap would be best inferred from the coincidence of higher energy interband transitions in the nodal $B2g$ spectra to those in the $B1g$ spectra.

\subsection{Density of States}
Given the difficulty of distinguishing $s$- and $d$-wave orders at low energy in a two-particle quantity, such as Raman spectroscopy, we instead compute the simplest measurable single particle quantity, the density of states, $N(\omega)=\sum\limits_{\vect{k}\sigma} A(\vect{k},\omega)$, which can be related to the tunnelling conductance, $dI/dV$ through
\begin{equation}
\frac{dI}{dV}=-\int d\omega f^\prime(\omega-{\rm V})N(\omega).
\end{equation}
The density of states has been examined several times using the YRZ ansatz with the inclusion of a d-wave pseudogap.\cite{borne:2010,storey:2013}  
Such a calculation can be fit to experimental results of the density of states which show a single asymmetric suppression of states near to the Fermi level.\cite{borne:2010, pushp:2009}  
Here we consider the impact of an $s$-wave pseudogap, which does not vanish at the node and is therefore expected to produce a reduction of $N(\omega)$ away from the Fermi level.

In the cuprates, STM has been most successful for observing the occurrence of spatial gaps near to the Fermi surface, identified by sharp quasiparticle peaks at $\pm \Delta$ centered at the origin of the superconducting gap, the Fermi level.  It is also well known that the inclusion of a bosonic density of states as is the case for phonon-mediated superconductivity creates an imprint of the bosonic couplings at energies away from the Fermi level.\cite{carbotte:2010}  Such couplings are thought to be responsible for peak-dip-hump structures seen in the cuprates.\cite{fischer:2007}  In this case, we will see the impact of a simplified hybridization coupling and show that there are relevant structures in $N(\omega)$ at frequencies away from the Fermi level, which are indeed sensitive to the structure of the pseudogap order.

\begin{figure}
\centering
  \includegraphics[width=.85\linewidth]{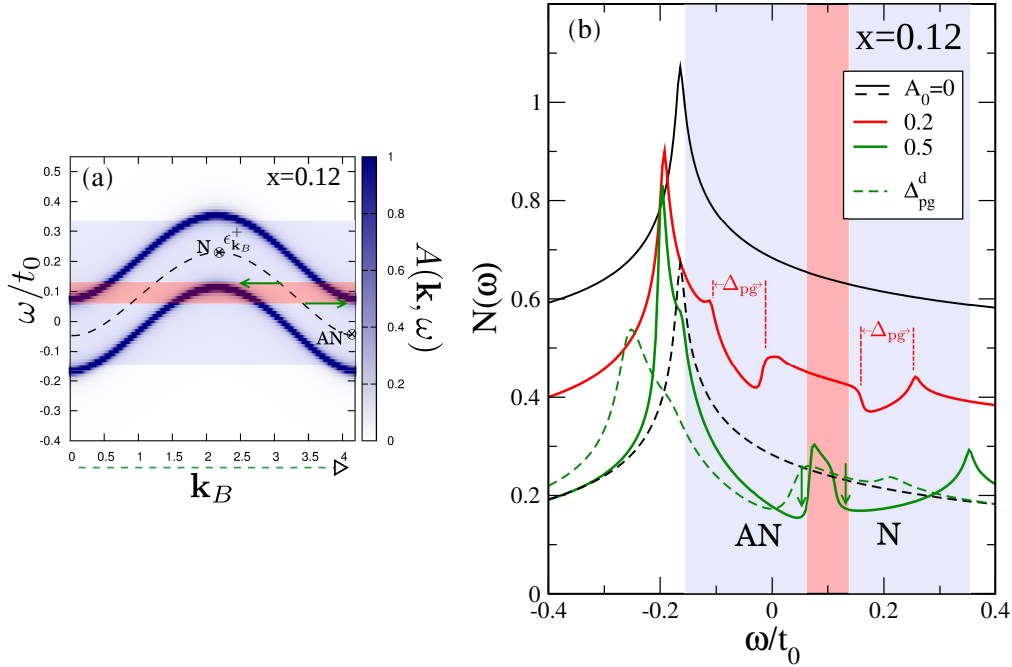}
    \caption{(Color Online) (a)  Reproduction of Fig.~\ref{fig:nacont}(b) with shading to identify regions of suppressed (blue) and enhanced (red) density of states.  (b)Normalized density of states, $N(\omega)$ for increasing $\Delta_{\rm pg}^s$ at fixed doping of $x=0.12$.  Marked with arrows are the suppression of $N(\omega)$ near the Fermi level due to a loss of antinodal (AN) states and a new feature at positive bias due to the loss of nodal (N) states at energies above the Fermi level. Each curve is offset by 0.2 for clarity.  The dashed black curve is $A_0=0$, with no offset to emphasize the regions of enhanced and suppressed $N(\omega)$. A single case for $\Delta_{\rm pg}^d$ at $A_0=0.5$ is shown (dashed-green).}
\label{fig:dos}
\end{figure}

We present in Fig.~\ref{fig:dos}(b) a representative case of the density of states for $x=0.12$ to illustrate the effects of a gradually increasing gap near the pseudogap onset.
  In our present formulation, the particle-hole  asymmetry results in two separate effects; a suppression of $N(\omega)$ near to zero bias, and a secondary depressed structure at positive bias.  Separating the two features is a hump structure caused by weight shifted from below (antinodal region) and above (nodal region) in energy.\cite{fischer:2007, gabovich:2013} 
In order to understand the necessity of two distinct gaps from a single order, we show for the $A_0=0.5$ case the spectral density, $A(\vect{k},\omega)$, as a function of $\vect{k}_B$ in Fig.~\ref{fig:dos}(a) using light-blue shading for the region of width $\Delta_{\rm pg}$ in both the nodal and anti-nodal sectors and light-red shading for the region in-between.  Considering these regions in Fig.~\ref{fig:dos}(b), the light-red region shows an increase in $N(\omega)$ as compared to the normal state, while the light-blue regions are suppressed as compared to the normal state.  This means that a single momentum independent order parameter has resulted in three separate energy features due to the particle-hole asymmetry of the antibonding dispersion.
We also provide a contrasting example with a $d-$wave order (dashed-green).  The primary distinction is that the density of states is no longer suppressed at positive energies, and instead has a single suppression near the Fermi level.  This is the most clear distinction between $s-$ and $d-$wave ordering from a momentum averaged quantity.
The information regarding the value of the pseudogap order parameter in the nodal region is located at finite energies, not near to the Fermi level, where STM experiments generally focus.

It is worth remarking that such a positive bias suppression may have been seen experimentally (See for example, Fig.3(e) in Ref.~\cite{hanaguri:2004}) in connection with static `checkerboard’ charge ordering.  Such charge ordering models can be phrased as a hybridization phenomenology with one or more scattering $Q$ vectors and if the spatial modulation is correctly accounted for, the resulting spectra exhibit both a zero bias and positive bias suppression separated by a hump, suggestive of an isotropic in momentum electronic gap with a strongly particle­-hole asymmetric dispersion.

\section{Conclusions}\label{sec:conc}
We have modified a successful ansatz to include a momentum independent hybridization gap and have shown how such a phenomenology results in a false $d$-wave pseudogap along the Fermi surface.   
We then compute the electronic Raman response at the onset of a pseudogap and identify the features of a momentum independent order.  From this we assert that a suppression of Raman signal at low energies does not indicate a nodal pseudogap, but is a generic sign of the loss of antinodal Fermi surface states which reduces the intraband Drude response.  A nodal pseudogap can only be suggested experimentally by an interband transition in $\chi_{B2g}^{\prime \prime}(\omega)$ which would correlate with a similar transition in the $\chi_{B1g}^{\prime \prime}(\omega)$ spectra.
Further, we identify a two gap behavior in the density of states which occurs as an artifact of the dispersive nature of the bonding band, $\epsilon_{\vect{k}}^+$, which has separate energies in the antinodal and nodal regions of the Brillouin zone.  This two scale behaviour in $N(\omega)$ is the clearest signature of a momentum independent pseudogap order.

   There exists now a significant literature comparing results of the YRZ ansatz to experimental work.\cite{rice:2012}
Previous works were focused on low energy probes in the presence of a d-wave superconducting gap which dominates low energy behaviour.\cite{illes:2009,leblanc:2009,borne:2010,james:2012, comin:2014}  We emphasize that in this work, the inclusion of a momentum independent pseudogap would not significantly impact the low energy nodal region where d-wave superconductivity dominates. This need not be the case at higher energies, as described in Fig.~\ref{fig:dos}.
 This is a main conclusion of this work, where we have demonstrated that the density of states at energies away from the Fermi level will be sensitive to the details of the zeros of $\epsilon_{\vect{k}}^-$ which occur at momenta $\vect{k}_B$.  This second gap at positive bias, and the resulting hump in between,  should be observable in tunnelling experiments which would support the assertion of a momentum independent pseudogap.

\appendix
\section{YRZ Phenomenology}\label{sec:app:yrz}
The energy dispersions and weighting factors contain doping dependent hopping parameters: $t(x)=g_{t}(x)t_{0}+3g_{s}(x)J\chi/8$, $t^{\prime
}(x)=g_{t}(x)t_{0}^{\prime }$, and $t^{\prime\prime
}(x)=g_{t}(x)t_{0}^{\prime \prime }$,  where $t_0$ is the bare hopping while $g_{t}(x)=\frac{2x}{1+x}$ and $g_{s}(x)=\frac{4}{(1+x)^{2}}$ are the energy renormalizing Gutzwiller factors for the kinetic and spin terms, respectively. \cite{vollhardt:1984,gutzwiller:1963}
 The renormalization factors provide an approximate dependence of the hopping parameters on the doping of the system which allows for a loss of metallicity as one approaches the Mott insulator transition at half filling ($x=0$). 
 Values of
other parameters in the dispersion are taken from Ref.~\cite{yrz:2006} to be: $t^{\prime}_0/t_{0}=-0.3$, $t^{\prime\prime}_0/t_{0}=0.2$,
$J/t_{0}=1/3$, and $\chi=0.338$ which are accepted values for the hole doped cuprates \cite{ogata:2008}. $\Delta_{\rm pg}^0(x)$  is then described phenomenologically in the doping phase diagram by the pseudogap line $\Delta_{pg}^0(x)= 3t_0(0.2-x)$.
We further enforce a Luttinger sum rule which is solved self-consistently for $\mu_p$ to obtain the desired hole doping, $x$, in the hybridized system.  It is important to note that in a fractionalized picture the Luttinger theorem is modified to account for states in each of the spinon/holon channels.  This has been addressed previously\cite{qi:2010} and  here we avoid such details and instead maintain the simplest Luttinger theorem for the ansatz of the hybridized $G(\vect{k},\omega, x)$.
This can be expressed as
\begin{equation}\label{eqn:lutt}
1-x=2\sum\limits_\vect{{\rm BZ}} \Theta\left( \Re e\left[ G(\vect{k},\omega=0, x)\right] \right)
\end{equation}
where $\Theta$ is the Heaviside function and $x$ is the percent of hole doping. 

\section{Computation of Raman Spectra}\label{sec:app:raman}
 The Raman response, $\chi_{\eta}^{\prime\prime}(\Omega)$, is given by
\begin{align}
\chi^{\prime \prime}_{\eta}(\Omega)=&\frac{\pi}{4}\sum_{\boldsymbol{k}}(\gamma_{\boldsymbol{k}}^\eta)^2\int_{-\infty}^{\infty}d\omega[f(\omega)-f(\omega+\Omega)]  \nonumber \\
&\times [A(\boldsymbol{k},\omega)A(\boldsymbol{k},\omega+\Omega)], \label{eq:raman} 
\end{align}
where $\eta$ is the choice of vertex B$_{1g}$ or B$_{2g}$ and $f(\omega)$ is the Fermi function, which we evaluate in the $T\to0$ limit. The vertex amplitude, $(\gamma^\eta_{\boldsymbol{k}})^2$, can be determined from the electronic energy dispersion,\cite{devereaux:2007} such that
\begin{align}
\gamma_{\boldsymbol{k}}^{B_{1g}}&=\frac{\partial^2 \epsilon_{\vect{\boldsymbol{k}}}^C}{\partial k_x^2}-\frac{\partial^2 \epsilon_{\vect{\boldsymbol{k}}}^C}{\partial k_y^2}, \\
\gamma_{\boldsymbol{k}}^{B_{2g}}&=\frac{\partial^2 \epsilon_{\vect{\boldsymbol{k}}}^C}{\partial k_x \partial k_y}.
\end{align}
Throughout this work we employ an elastic scattering rate of $\Gamma=0.04 t_0$ to give width to the low energy Drude response.

\ack
JPFL would like to thank Peter Fulde and Andrey Chubukov for useful discussion leading to this work.
\vspace{12pt}

\bibliographystyle{new2}
\bibliography{bib-test2}

\end{document}